\documentclass[11pt,a4paper]{article}
\usepackage{jheppub}
\usepackage{pdflscape}
\usepackage{amsmath}
\usepackage{amssymb}
\usepackage{dcolumn}
\usepackage{bm}
\usepackage{color}
\usepackage{epsfig}
\usepackage{amsfonts}
\usepackage{graphicx}
\usepackage{subfigure}
\usepackage{dcolumn}

\newcommand{\be}{\begin{equation}}
\newcommand{\ee}{\end{equation}}
\newcommand{\bea}{\begin{eqnarray}}
\newcommand{\eea}{\end{eqnarray}}

\def\be{\begin{equation}}
\def\ee{\end{equation}}
\def\bea{\begin{eqnarray}}
\def\eea{\end{eqnarray}}

\begin{document}

\title{Noether Symmetry Approach in Eddington-inspired Born-Infeld gravity}

\author[a]{Thanyagamon Kanesom,}
\author[b,c,d]{Phongpichit Channuie,}
\author[a]{Narakorn Kaewkhao}

\affiliation[a] {Department of Physics, Faculty of Science, Prince of Songkla University, Hatyai 90112, Thailand}
\affiliation[b] {School of Science, Walailak University, Nakhon Si Thammarat, 80160, Thailand} 
\affiliation[c] {College of Graduate Studies, Walailak University, Nakhon Si Thammarat, 80160, Thailand}
\affiliation[d] {Research Group in Applied, Computational and Theoretical Science (ACTS), \\Walailak University, Nakhon Si Thammarat, 80160, Thailand} 

\emailAdd{thanyagamon1995@gmail.com}
\emailAdd{channuie@gmail.com}
\emailAdd{naragorn.k@psu.ac.th}

\abstract{In this work, we take a short recap of a formal framework of the Eddington-inspired Born-Infeld (EiBI) theory of gravity and derive the point-like Lagrangian for underlying theory based on the use of Noether gauge symmetries (NGS). We study a Hessian matrix and quantify Euler-Lagrange equations of EiBI universe. We discuss the NGS approach for the Eddington-inspired Born-Infeld theory and show that there exists the de Sitter solution in this gravity model.}


\maketitle

%%%%%%%%%%%%%%%%%%%%%%%%%%%%
\section{Introduction}
%%%%%%%%%%%%%%%%%%%%%%%%%%%%
Various cosmological observations make a strong evidence that the expansion of the universe is presently accelerating. These experimental results include Type Ia Supernovae \cite{Perlmutter:1998np,Riess:1998cb}, 
cosmic microwave background (CMB) radiation \cite{Ade:2015xua, Ade:2015lrj, Ade:2014xna, Ade:2015tva, Array:2015xqh, Komatsu:2010fb, Hinshaw:2012aka}, 
large scale structure \cite{LSS}, 
baryon acoustic oscillations (BAO) \cite{Eisenstein:2005su} 
as well as weak lensing \cite{Jain:2003tba}. 
An expansion phase can be basically explained by the simplest model: the so-called Lambda cold dark matter (${\Lambda}{\rm CDM}$) \cite{Akrami:2018vks}. However, the ${\Lambda}{\rm CDM}$ model is plagued by the cosmological problem \cite{Weinberg:2000yb} and the coincident problem \cite{Velten:2014nra}. There are at least two promising explanations to date to describe the late-time cosmic acceleration. The first one  assumes the introduction of the so-called ``dark energy (DE)'' in the context of conventional general relativity. Another convincing approach is to engineer Einstein gravity on the large-scale methodology (see for reviews on not only dark energy problem but also modified gravity theories, e.g., \cite{Nojiri:2010wj,Nojiri:2006ri,Book-Capozziello-Faraoni,Capozziello:2011et,Bamba:2015uma}). However, the DE sector remains still unknown and possesses one of the unsolved problems in physics.

Therefore, it opens opportunities to search for modified theories of gravity to deal with such problems. By modifying the geometrical part of Einstein field equations or adding  scalar field to the right-hand side of the Einstein field equations, both alternatives are able to explain effects of dark ingredients with acceptable assertions \cite{Copeland:2006wr}. One of the simplest modifications to the standard general relativity is the $f(R)$ theories of gravity
in which the Lagrangian density $f$ is an arbitrary function of the scalar curvature $R$ \cite{Bergmann:1968ve,Buchdahl:1983zz}. Among
numerous alternatives, these theories include higher order curvature invariants, see rigorous reviews on $f(R)$ theories \cite{Sotiriou:2008rp,DeFelice:2010aj}.

In cosmological framework, the Noether symmetry (NS) approach has revealed a useful tool not only to fix physically viable cosmological models with respect to the conserved quantities, but also to reduce dynamics and achieve exact solutions \cite{Copoz_a,Copaz_b}. Moreover, the existence of Noether symetries plays crucial roles when studying quantum cosmology \cite{SC}. The Noether symmetry approach has been employed to various cosmological scenarios so far including the $f(T)$ gravity \cite{Channuie:2017txg}, the $f(R)$ gravity \cite{j}, the alpha-attractors \cite{Kaewkhao:2017evn}, spherical and cylindrical solutions in $f(T)$ gravity \cite{Nurbaki:2020dgw}, $f(G)$\,gravity \cite{Bajardi:2020osh}, non-local curvature and Gauss–Bonnet cosmologies \cite{Bajardi:2020mdp}, and others cosmological scenarios, e.g. \cite{cap3,Momeni:2015gka,Momeni:2014iua,Aslam:2012tj,Jamil:2012fs,Jamil:2011pv,Bahamonde:2017sdo}. The study of Palatini $f(R)$ cosmology using the NS approach for the matter-dominated universe was carried out in Ref.\cite{Capozziello,camci}. Moreover, the exact solutions for potential functions, scalar field and the scale factors in the Bianchi models have been investigated in \cite{Jamil:2012zm,Channuie:2018now}.

Apart from the NS approach \cite{Copoz_a,Copaz_b}, the Noether Gauge Symmetry (NGS) \cite{Aslam:2013pga,Hussain:2011wa,Kucuakca:2011np} is more generalized. In this work, we examine a formal framework of Eddington-inspired Born-Infeld (EiBI) gravity through the NGS approach and present a detailed calculation of the point-like Lagrangian. Notice that the point-like Lagrangian derived from the alternative form of the EiBI action was proposed by Delsate and Steindoff \cite{Delsate:2012ky} instead of using the original form of the EiBI action suggested by M$\rm\acute{a}$ximo Ba$\rm\tilde{n}$adoz \cite{Banados_progony:2009}.

%%% Organization %%%
This paper is organized as follows: We will start by making a short recap of a formal framework of the Eddington-inspired Born-Infeld theory of gravity in Sec.\ref{EiBI}. Here we derive the point-like Lagrangian for underlying theory. In Sec.\ref{Hess}, we study a Hessian matrix and quantify Euler-Lagrange equations of EiBI universe. In Sec.\ref{NGS}, the NGS approach for the Eddington-inspired Born-Infeld theory is discussed. We comment on exact cosmological solutions of the EiBI theory based on the use of Noether symmetries of point-like Lagrangians in Sec.\ref{Sols}. Finally, we conclude our findings in the last section.

%%%%%%%%%%%%%%%%%%%%%%%%%%%
\section{Eddington-inspired Born-Infeld Gravity}\label{EiBI}
%%%%%%%%%%%%%%%%%%%%%%%%%%%
In 2009, M$\rm\acute{a}$ximo Ba$\rm\tilde{n}$adoz \cite{Banados_progony:2009} proposed a new form of the Born-Infeld action under Palatini formalism. This is the so-called Eddington-inspired-Born-Infeld (EiBI) gravity.  This action is written as follows:
\bea\label{EiBI v2}
S_{\rm EiBI}(g,\Gamma)=\frac{2}{\kappa}\int d^4x \Big[\sqrt{|g_{\mu\nu}+\kappa R_{\mu\nu}(\Gamma)|}-\lambda\sqrt{\,|g_{\mu\nu}|}\,\Big] +S_{m}(g_{\mu\nu},\Psi)+S_{\phi}(g_{\mu\nu},\phi),
\eea
where $\lambda=1+\kappa\Lambda$ is a dimensionless constant displaying the relation between the EiBI free parameter $\kappa$ (with a dimension of $M^{2}_{P}$ ) and the cosmological constant $\Lambda$ (with a dimension of $M^{-2}_{P}$); whilst $S_{m}(g_{\mu\nu},\Psi)$ and $S_{\phi}(g_{\mu\nu},\phi)$ represent the matter field action and the scalar field action, respectively.  Throughout this work, we set $\frac{8\pi G_{\rm N}}{c^{4}}=1.$ 
Performing variation of Eq.(\ref{EiBI v2}) with respect to $\Gamma^{\lambda}_{\mu\nu}$ gives the relation between two metric tensors, i.e. $q_{\mu\nu} = g_{\mu\nu}+\kappa R_{\mu\nu}(\Gamma).$ Hence equation (\ref{EiBI v2}) can be written in the bi-metric form as
\bea\label{EiBI v2 rewritten}
S_{\rm EiBI}(g,\Gamma)=\frac{2}{\kappa}\int d^4x \Bigg[\sqrt{|q_{\mu\nu}|}-\lambda\sqrt{\,|g_{\mu\nu}|}\,\Bigg] +S_{m}(g_{\mu\nu},\Psi)+S_{\phi}(g_{\mu\nu},\phi).
\eea
With the help of two ansatz forms of a spatially flat FLRW metric
\bea\label{metric g}
ds^{2}_{g} &=& g_{\mu\nu}dx^{\mu}dx^{\nu}= -N(t)^{2}dt^{2}+a(t)^{2}d{\vec{x}}^{2},\\ 
ds^{2}_{q}&=& q_{\mu\nu}dx^{\mu}dx^{\nu}=-M(t)^{2}dt^{2}+b(t)^{2}d{\vec{x}}^{2},
\eea 
the EiBI action (\ref{EiBI v2 rewritten}) can be expressed in terms of the cosmological variables as follows:
\begin{eqnarray}\label{EiBI full point-like}
S_{\rm EiBI}&=&\frac{2}{ \kappa}\upsilon_{0}\int dt \mathcal{L}_{\rm EiBI}\nonumber\\&=&\frac{2\upsilon_{0}}{ \kappa}\int dt \Bigg[\sqrt { a^{6}\Big[N^{2}-3\kappa (\frac{\dot{M}\dot{b}}{Mb}-\frac{\ddot{b}}{b})\Big]\Big[1+\frac{\kappa}{a^{2}M^{2}}(-\ddot{b}b+2\dot{b}^{2}-\frac{b\dot{b}\dot{M}}{M}) \Big]^{3}}\nonumber\\ &&-Na^{3}(1+\kappa \Lambda)\Bigg]-Na^{3}\Big[ -\rho_{m}(a)+\frac{l\dot{\phi}^{2}}{N^{2}}-2V(\phi)  \Big],
\end{eqnarray}
where $\upsilon_{0}$ is the spatial volume obtaining after a proper compactification for spatial flat section. $l=+1$ and $l=-1$ denote an ordinary scalar field and phantom scalar field, respectively. It is probably impossible to get rid of all second order derivative in Eq.(\ref{EiBI full point-like}) by performing an integration by parts. Therefore we neglect to write an explicit form of the point-like Lagrangian. In 2012, Delsate and Steindoff \cite{Delsate:2012ky}, however, wisely proposed the bi-metric form of the EiBI action under the metric formulation written as
\begin{eqnarray}\label{Delsate EiBI}
S_{\rm EiBI}(g,q) &=&\lambda \int d^{4}x \sqrt{-q}\Bigg[R(q)-\frac{2\lambda}{\kappa}+\frac{1}{\kappa} \Bigg(q^{\alpha\beta} g_{\alpha\beta} -2\sqrt{\frac{g}{q}}\Bigg)\Bigg]\nonumber\\&&+S_{\phi}(g,\phi)+S_{M}(g,\Phi).
\end{eqnarray}
Notice that only $g_{\mu\nu}$ interacts with matter and scalar fields whereas $q_{\mu\nu}$ defines the background metric regarded as the fundamental reference frame of the universe \cite{Rosen-bi}. 
Because the alternative action yields identical field equations as provided in the Ba$\rm\tilde{n}$adoz action, this indicates that two action forms are equivalent. Using  a relation, $S=\nu_{0} \int{dt}\mathcal{L}$ \cite{Bouhmadi-Lopez:2016dcf} and the relation between two  metric tensors, $q_{\mu\nu} = g_{\mu\nu}+\kappa R_{\mu\nu}(\Gamma)$, the point-like Lagrangian  can be extracted from equation(\ref{Delsate EiBI}) as follows
\begin{equation} \label{Point-like Lagrangian EiBI 1}
\mathcal{L}_{\rm EiBI}=\lambda M b^{3}\Bigg{[}-\frac{6{\dot{{b}}}^{2}}{M^{2} b^{2}} -\frac{2\lambda}{\kappa} +\frac{1}{\kappa} \Bigg{(} \frac{N^{2}}{M^{2}} +3\frac{a^{2}}{b^{2}}\Bigg{)}\Bigg{]}+Na^{3} \Bigg{(}l\frac{{\dot{\phi}}^{2}}{N^{2}}-2V(\phi)-2\rho_{m}(a)\Bigg{)}.
\end{equation}
Here the number of configuration space (or the minisuperspace) variables equal to five due to the appearance of variables $\{a(t),b(t),M(t),N(t),\phi(t)\}$ in Eq.(\ref{Point-like Lagrangian EiBI 1}). Apart from the kinetic part of the Lagrangian, we can set $V_{\rm eff}=\frac{\lambda M b^{3}}{\kappa}\Big{[} {2\lambda} - \Big{(} \frac{N^{2}}{M^{2}} +3\frac{a^{2}}{b^{2}}\Big{)}\Big{]}+2Na^{3} \Big{(}V(\phi)+\rho_{m}(a)\Big{)}$ as an effective potential in the gravity model. It also notes that Eq.(\ref{Point-like Lagrangian EiBI 1}) is a singular Lagrangian due to the existence of two Lapse functions, $N(t)$ and $M(t)$ as shown in the denominators of Eq.(\ref{Point-like Lagrangian EiBI 1}).
%%%%%%%%%%%%%%%%%%%%%%%%%%%%%%%%%%%%
\section{\textbf{Hessian matrix and Euler-Lagrange equations of EiBI universe }}
\label{Hess}
%%%%%%%%%%%%%%%%%%%%%%%%%%%%
%The affine parameter here is the cosmic time in cosmological context. From Eq.(\ref{EIBI PL}),
%the configuration space variables and their time derivative in EiBI point-like Lagrangian are $q^{i}= \{a,b,N,M,\phi\}$  and $\dot{q}^{i}=\frac{dq_{i}}{dt} =\{\dot{a},\dot{b},\dot{N},\dot{M},\dot{\phi}\}$ respectively.
In the absent of $\{\dot{a},\dot{M},\dot{N}\} $ in the EiBI point-like Lagrangian, the EiBI Hessian matrix can be written as
%are not equal to zero, i.e. $\frac{\partial^{2}{L}}{\partial \dot{b}^{2}}=-\frac{12\lambda b}{M} $ and $\frac{\partial^{2}{L}}{\partial \dot{\phi}^{2}}=\frac{2a^{3}l}{N}.$  \\
\begin{eqnarray}
[W_{ij}]_{\rm EiBI}=\left[
                       \begin{array}{ccccc}
                         \frac{\partial^{2}L}{\partial\dot{a}^{2}} &\frac{\partial^{2}L}{\partial\dot{a}\partial \dot{b}} & \frac{\partial^{2}L}{\partial\dot{a}\partial \dot{N}} & \frac{\partial^{2}L}{\partial \dot{a}\partial\dot{M}} & \frac{\partial^{2}L}{\partial\dot{a}\partial \dot{\phi}} \\
                         \frac{\partial^{2}L}{\partial\dot{b}\partial \dot{a}} & \frac{\partial^{2}L}{\partial\dot{b}^{2}} & \frac{\partial^{2}{L}}{\partial\dot{b}\partial \dot{N}} &\frac{\partial^{2}{L}}{\partial \dot{b}\partial\dot{M}}&\frac{\partial^{2}{L}}{\partial\dot{b}\partial \dot{\phi}}\\
                         \frac{\partial^{2}L}{\partial\dot{N}\partial \dot{a}} & \frac{\partial^{2}L}{\partial \dot{N}\partial\dot{b}} & \frac{\partial^{2}L}{\partial\dot{N}^{2}} &\frac{\partial^{2}{L}}{\partial\dot{N}\partial \dot{M}} &\frac{\partial^{2}{L}}{\partial \dot{N}\partial\dot{\phi}} \\
                         \frac{\partial^{2}L}{\partial\dot{M}\partial \dot{a}} & \frac{\partial^{2}L}{\partial \dot{M}\partial\dot{b}} & \frac{\partial^{2}L}{\partial\dot{M}\partial\dot{N}} &\frac{\partial^{2}L}{\partial\dot{M}^{2}} &\frac{\partial^{2}L}{\partial\dot{M}\partial \dot{\phi}} \\
                         \frac{\partial^{2}L}{\partial\dot{\phi}\partial\dot{a}} & \frac{\partial^{2}L}{\partial \dot{\phi}\partial \dot{b}} & \frac{\partial^{2}L}{\partial\dot{\phi}\partial\dot{N}} &\frac{\partial^{2}L}{\partial\dot{\phi}\partial\dot{M}} &\frac{\partial^{2}L}{\partial \dot{\phi}^{2}}
                       \end{array}
                     \right]=
                    \left[
                       \begin{array}{ccccc} 
                   0 & 0 & 0 & 0 & 0  \\  
                   0 & -\frac{12\lambda b}{M} & 0 & 0 & 0\\
                   0 & 0 & 0 & 0 & 0\\
                   0 & 0 & 0 & 0 & 0\\
                   0 & 0 & 0 & 0 & \frac{2a^{3}l}{N}
                        \end{array}
                     \right].
\end{eqnarray}
Clearly, the determinant of the Hessian matrix of the EiBI point-like Lagrangian equals zero indicates again that Eq.(\ref{Point-like Lagrangian EiBI 1}) is a singular Lagrangian. Accordingly, variables $\{a,N,M\} $ do not contribute to dynamics and have to be considered as a further constraint equations. This tells us that $a(t),M(t)$ and $N(t)$ are not independent variables anymore then we  can set them to an arbitrary functions of time\cite{MartinB}, i.e. $\mathcal{F}(t)=\mathcal{F}(a,M,N)$. Variables $b(t)$ and $\phi(t)$ however remain considered independently. With the definition of the Euler-Lagrange equations, we can show that \cite{Henneaux:1992ig}
\bea\label{degenete anm1}
\frac{d}{dt}\frac{\partial \mathcal{L}}{\partial\dot{ q}^{i}}-\frac{\partial \mathcal{L}}{\partial q^{i}}& = &
\frac{\partial}{\partial q^{j}}\Big(\frac{\partial \mathcal{L}}{\partial \dot{q}^{i}}\Big)\frac{dq^{j}}{dt}+\frac{\partial}{\partial \dot{q}^{j}}\Big(\frac{\partial \mathcal{L}}{\partial \dot{q}^{i}}\Big)\frac{d\dot{q}^{j}}{dt}-\frac{\partial \mathcal{L}}{\partial q^{i}},\nonumber\\
%=\frac{\partial^{2}\mathcal{L}}{\partial q^{j}\partial \dot{q}^{i}}\dot{q}^{j}+\frac{\partial^{2}\mathcal{L}}{\partial \dot{q}^{i}\partial \dot{q}^{j}}\ddot{q}^{j}-\frac{\partial \mathcal{L}}{\partial q^{i}}\no \\&=&
&=&\ddot{q}^{j}\frac{\partial^{2} \mathcal{L}}{\partial \dot{q}^{j}\partial \dot{q}^{i}}+\dot{q}^{j}\frac{\partial^{2} \mathcal{L}}{\partial q^{j}\partial \dot{q}^{i}}-\frac{\partial \mathcal{L}}{\partial q^{i}}=0,\\
\ddot{q}^{j}\frac{\partial^{2} \mathcal{L}}{\partial \dot{q}^{j}\partial \dot{q}^{i}}   &=&  -\dot{q}^{j}\frac{\partial^{2} \mathcal{L}}{\partial q^{j}\partial \dot{q}^{i}}+\frac{\partial \mathcal{L}}{\partial q^{i}}.\label{RHs anm1}
\eea
Here the configuration space variables are $q_{i}= \{a,b,M,N,\phi\}$ and their time derivative on the tangent space are $\dot{q}_{i}=\frac{dq_{i}}{dt} =\{\dot{a},\dot{b},\dot{M},\dot{N},\dot{\phi}\}$.
Because $\frac{\partial^{2} \mathcal{L}}{\partial \dot{q}^{j}\partial \dot{q}^{i} }= 0$ and $\frac{\partial^{2} \mathcal{L}}{\partial q^{j}\partial \dot{q}^{i}}$   for variables $\{ a, M,N \}$ in EiBI gravity, the Euler-Lagrange equations of these variables can be reduced to $\frac{\partial \mathcal{L}}{\partial a}=\frac{\partial \mathcal{L}}{\partial M}=\frac{\partial \mathcal{L}}{\partial N}=0$ where  $\dot{a},\dot{M},\dot{N}$ and  $\ddot{a}, \ddot{M}, \ddot{N}$ can be set arbitrarily \cite{Henneaux:1992ig}.
% hence $\ddot{a},\ddot{N}, \ddot{M}$ can be set arbitrarily.
As expected, $\ddot{b}$ and $\ddot{\phi}$ are determined  by performing variation equation(\ref{Point-like Lagrangian EiBI 1}) with respect to the dynamical variables and their time derivative as expressed on the right-hand side of equation(\ref{degenete anm1}).
As expected, $\ddot{b}$ and $\ddot{\phi}$ are determined  from taking variation Lagrangian with respect to the dynamical variables and their time derivative as shown on the right-hand side of Eq.(\ref{RHs anm1}). We have to keep in mind that a crucial concept of a gauge theory is the general solution of the equations of motion which contains arbitrary functions of time and the canonical variables are not all independent but relate among each others via the constraint equations \cite{Henneaux:1992ig}. As the results of the vanishing of $\frac{\partial^{2} \mathcal{L}}{\partial \dot{a}\partial \dot{a}}\equiv\frac{\partial^{2} \mathcal{L}}{\partial \dot{a}^{2}},\frac{\partial^{2} \mathcal{L}}{\partial \dot{M}^{2}}$ and $\frac{\partial^{2} \mathcal{L}}{\partial \dot{N}^{2}}$, the canonical momenta associated to $a,N,$ and $M$  yield $p_{a}={\partial\mathcal{ L}}/{\partial \dot{a}}=0, p_{N}={\partial\mathcal{ L}}/{\partial \dot{N}}=0, p_{M}={\partial\mathcal{ L}}/{\partial \dot{M}}=0$, respectively.
%, these imply the dynamics of the non-dynamical degree of freedom\cite{Remmen:2013eja}, i.e. $\{ a,N,M \} $ are not independent variables anymore (see Ref.\cite{Sidney colemann}, section 4.1).
%This implies that $a(t),N(t)$ and $M(t)$ are arbitrary functions of time \cite{MartinB} and are not independent variables anymore (see ref.\cite{Sidney colemann}, section 4.1).
The Hamiltonian constraint equation can be straightforwardly derived from the canonical momenta via the Lagrangian and theirs Lagrange multipliers $\{\lambda_{i}=\lambda_{a}(t),\lambda_{M}(t),\lambda_{N}(t)\}$ as follows:
\begin{eqnarray}\label{Hamiltonian constraint}
\mathcal{H}_{\rm EiBI}&=&\frac{\partial{{\mathcal{L}}_{\rm EiBI}}}{\partial \dot{q_i}}\dot{q_i}-{\mathcal{L}}_{\rm EiBI}
=\frac{\partial {\mathcal{L}}_{\rm EiBI}}{\partial \dot{b}}\dot{b}+\frac{\partial {\mathcal{L}}_{\rm EiBI}}{\partial \dot{\phi}}\dot{\phi}-{{\mathcal{L}}_{\rm EiBI}},
\end{eqnarray}
\begin{eqnarray}\label{Hamiltonian constraint1}
\mathcal{H}_{\rm EiBI,tot}&=&\mathcal{H}_{\rm EiBI}+\Sigma\lambda_{i}p_{i},\nonumber \\
&=&\Big(-\frac{12\lambda b\dot{b}}{M}\Big)\dot{b}+\Big(\frac{2a^{3}l \dot{\phi}}{N} \Big)\dot{\phi}-{{\mathcal{L}}_{\rm EiBI}}+\lambda_{a}p_{a}+\lambda_{N}p_{N}+\lambda_{M}p_{M}=0, \nonumber\\
%&=&\Big[-\frac{6\lambda b \dot{b}^{2}}{M}  +\frac{a^{3}l\dot{\phi}^{2}}{N}\Big]+\Big[ \frac{2\lambda^{2}Mb^{3}}{\kappa}-\frac{\lambda b^{3}}{\kappa}
%(\frac{N^{2}}{M}+\frac{3a^{2}M}{b^{2}})+2Na^{3}V(\phi) +2\rho(a)Na^3  \Big], \no\\
&=& \Big[-\frac{6\lambda b \dot{b}^{2}}{M}  +\frac{a^{3}l\dot{\phi}^{2}}{N}\Big]+\frac{1}{\kappa}\Big[ {2\lambda^{2}Mb^{3}}-{\lambda b^{3}}
\frac{N^{2}}{M}-{3\lambda a^{2}Mb}\nonumber\\
&&+2\kappa Na^{3}V(\phi) +2\kappa\rho(a)Na^3  \Big]+\lambda_{a}p_{a}+\lambda_{M}p_{M}+\lambda_{N}p_{N}=0,
\end{eqnarray}
where $\lambda_{a}(t),\lambda_{M}(t)$,and $\lambda_{N}(t)$ stand for Lagrange multipliers that are arbitrary functions of time.
The total EiBI Hamiltonian $(\mathcal{H}_{\rm EiBI})$ can be used to evaluate an evolution invoking the Hamiltonian equations of motion as follows:
\begin{eqnarray}\label{Hamiltonian eqs EiBI}
\dot{a}&=&\frac{\partial\mathcal{H}_{\rm EiBI,tot} }{\partial p_{a}}=\lambda_{a}(t),\,
\dot{M}=\frac{\partial\mathcal{H}_{\rm EiBI,tot} }{\partial p_{M}}=\lambda_{M}(t),\,
\dot{N}=\frac{\partial\mathcal{H}_{\rm EiBI,tot} }{\partial p_{N}}=\lambda_{N}(t), \nonumber\\
\dot{b}&=&\frac{\partial\mathcal{H}_{\rm EiBI,tot} }{\partial p_{b}}=-\frac{p_{b}M}{12\lambda b},\,\,
\dot{\phi}=\frac{\partial\mathcal{H}_{\rm EiBI,tot} }{\partial p_{\phi}}=\frac{p_{\phi}N}{2a^{3}l},
\nonumber\\
\dot{p}_{b}&=&-\frac{\partial\mathcal{H}_{\rm EiBI,tot} }{\partial b}=\frac{6\lambda \dot{b}^{2}}{M}-\frac{1}{\kappa}\Big[6\lambda^{2}Mb^{2}-\frac{3\lambda b^{2}N^{2}}{M} -3\lambda a^{2}M  \Big]\neq0,\nonumber\\
 \dot{p}_{\phi}&=&-\frac{\partial\mathcal{H}_{\rm EiBI,tot} }{\partial \phi}=-\frac{1}{\kappa}\Big[ 2\kappa Na^{3}V'(\phi)\Big]=-2Na^{3}V'(\phi)\neq0,\nonumber \\
 \dot{p}_{a}
&=&-\frac{\partial\mathcal{H}_{\rm EiBI,tot} }{\partial a}= \frac{3a^{2}l\dot{\phi}^{2}}{N}+\frac{1}{\kappa}\Big[ 6\lambda abM-6\kappa a^{2} NV(\phi)+6\kappa a^{2}p_{m}(a)N \Big]=0,\nonumber
\\
\dot{p}_{N}&=&-\frac{\partial\mathcal{H}_{\rm EiBI,tot} }{\partial N}=-\frac{a^{3}l\dot{\phi}^{2}}{N^{2}}+\frac{1}{\kappa}\Big[ \frac{2\lambda b^{3}N}{M}-2\kappa a^{3}V(\phi)-2\kappa \rho_{m}(a)a^{3} \Big]=0,\nonumber\\
 \dot{p}_{M}&=&-\frac{\partial\mathcal{H}_{\rm EiBI,tot} }{\partial M}=-\frac{6\lambda b\dot{b}^{2}}{M^{2}}+\frac{1}{\kappa}\Big[ -2\lambda^{2}b^{3}-\frac{\lambda b^{3} N^{2}}{M^{2}}+3\lambda a^{2}b  \Big]=0.
\end{eqnarray}
For the EiBI Hamiltonian, it should be noted that $\dot{a}=\lambda_{a}, \dot{N}=\lambda_{N}, \dot{M}=\lambda_{M}$ are the primary constraints and $\dot{p}_{a}=\dot{p}_{N}=\dot{p}_{M}=0$ are the secondary constraints that must be valid at all times \cite{MartinB} resulting in $p_{a}=p_{M}=p_{N}=0$.
In order to obtain the dynamic solutions, we have to calculate the Euler-Lagrange equations for $a(t),b(t),M(t), N(t)$, and $\phi(t)$ as shown below:
\begin{eqnarray}\label{EL EiBI}
\frac{3a^{2}l\dot{\phi}^{2}}{N}+\frac{6\lambda ab  M}{\kappa}-6Na^{2}V(\phi)+6Na^{2}p_{m}(a)&=&0,\label{EL a}\\
\ddot{b}-\frac{\dot{b}\dot{M}}{M}+\frac{\dot{b}^{2}}{2b}-\frac{\lambda M^{2}b}{2\kappa}+\frac{bN^{2}}{4\kappa}+\frac{a^{2}M^{2}}{4\kappa b}&=&0,
\label{dd B full2}\\
\frac{6\lambda b\dot{b}^{2}}{M^{2}}-\frac{2\lambda^{2} b^{3}}{\kappa}+\frac{3\lambda a^{2} b}{\kappa}-\frac{\lambda b^{3}}{\kappa}\frac{N^{2}}{M^{2}}&=&0,\label{dd M}\\
-\frac{a^{3}l\dot{\phi}^{2}}{N^{2}}+\frac{2\lambda b^{3} N}{\kappa M}-2a^{3}V(\phi)-2a^{3}\rho_{m}(a)&=&0,\label{dd N}\\
\ddot{\phi}+\Big( 3\frac{\dot{a}}{a}+\frac{\dot{N}}{N}\Big)\dot{\phi}+\frac{V'(\phi)N^{2}}{l}&=&0,\label{dd phi2}
\end{eqnarray}
where the conservation equation \cite{Bouhmadi-Lopez:2016dcf}, $d\rho_{m}/da=-3(\rho_{m}+p_{m})/a,$ has been used to yield Eq.(\ref{EL a}).
\section{Noether Gauge Symmetries in EiBI Gravity}
\label{NGS}
%%%%%%%%%%%%%%%%%%%%%%%%%%%%%%%%%%%%%%%%%%%%%%%%%%%

%The approach of Noether gauge symmetry was applied to Eq.(\ref{Point-like Lagrangian EiBI 1}) aimed to specify cosmological functions of EiBI gravity.
Noether vector $(\mathrm{X}_{NGS})$ and the first prolongation vector field $(\mathrm{X}^{[1]}_{\rm NGS})$ related to EiBI Lagrangian, as shown in Eq.(\ref{Point-like Lagrangian EiBI 1}), can be constructed as follows:
\begin{equation}\label{vector field NGS}
\mathrm{X}_{NGS}=\tau\frac{\partial}{\partial t}+\alpha\frac{\partial}{\partial a}+\beta\frac{\partial}{\partial b}
+\gamma\frac{\partial}{\partial N}+\xi\frac{\partial}{\partial M}+\varphi\frac{\partial}{\partial \phi},
\end{equation}
\begin{eqnarray}\label{first prolongation}
 \mathrm{X}^{[1]}_{\rm NGS}&=&\mathrm{X}_{\rm NGS}+\dot{\alpha}\frac{\partial}{\partial \dot{a}}+ \dot{\beta}\frac{\partial}{\partial \dot{b}}+ \dot{\gamma}\frac{\partial}{\partial \dot{N}}
 +\dot{\xi}\frac{\partial}{\partial \dot{M}}+\dot{\varphi}\frac{\partial}{\partial \dot{\phi}},
\end{eqnarray}
where the undetermined  parameters $\{\tau(t,q^{i}),\alpha(t,q^{i}),\beta(t,q^{i}),\gamma(t,q^{i}),\xi(t,q^{i}),\varphi(t,q^{i})\}$ are possibly functioned by $\{t,q^{i}\}$=$\{t,a,b,N,M,\phi\}$. Their time derivative can be defined as
\begin{eqnarray}\label{dot alphasa}
\dot{\alpha}(t,a,b,N,M,\phi)=\mathrm{D}_{t}\alpha-\dot{a}\mathrm{D}_{t}\tau,
\end{eqnarray}
for variable $\alpha(t)$. This can be applied in the same way for other undetermined variables. The operator of a total differentiation ($\mathrm{D}_t$) with respect to $t$ in EiBI gravity can be defined as
\begin{equation}
    \mathrm{D}_t =\frac{\partial}{\partial t}+\dot{a}\frac{\partial}{\partial a}+ \dot{b}\frac{\partial}{\partial b}+ \dot{M}\frac{\partial}{\partial M}+ \dot{N}\frac{\partial}{\partial N}+ \dot{\phi}\frac{\partial}{\partial \phi}.
\end{equation}
It is worth noting that terms like, for example, $\frac{\partial{V(\phi(t))}}{\partial t}=\frac{\partial \dot{a}(t)^{2}}{\partial t}=0,$ because there is not time variable(t) shown explicitly in $\mathcal{L}_{\rm EiBI}.$ 
%Namely,$\frac{\partial\mathcal{L}[(q^{i}(t),{\dot{q}^{i}(t)]}}{\partial t}=0.$
The vector field $\mathrm{X}^{[1]}_{NGS}$  is a NGS of a Lagrangian $\mathcal{L}(t,a,b,\phi,M,N,\dot{b},\dot{\phi})$, if there exists a gauge function $\mathrm{B}(t, a, b,\phi, M, N)$ which obeys the following condition (see Ref.\cite{Kurt-Fund} for explicit derivation):
\begin{eqnarray}\label{NGS gaugcd1}
    \mathrm{X}^{[1]}_{NGS}\mathcal{L}+\mathcal{L}\mathrm{D}_t\tau=\mathrm{D}_t \mathrm{B}.
\end{eqnarray}
For NSG without gauge term and the prolongation part of the vector field, i.e. $\mathrm{B}(t,q^{i})=0$, it requires that $\tau(t,q^{i})=0.$ Accordingly, Eq.(\ref{NGS gaugcd1}) can be reduced to $\pounds_{X_{\rm NGS}}\mathcal{L}=0$ that is the condition for Noether symmetry \cite{Kucuakca:2011np}. After using the Noether gauge symmetries condition with the EiBI point-like Lagrangian, this provides us with eighty terms for $\mathrm{X}^{[1]}_{NGS}\mathcal{L}+\mathcal{L}\mathrm{D}_t\tau=\mathrm{D}_t \mathrm{B}$ as shown below:
\begin{eqnarray}
&&\frac{6\lambda a b M\alpha}{\kappa}-\frac{6\lambda a^2 N\alpha}{\kappa}-6a^2 N\rho(a)\alpha-6a^2 NV(\phi)\alpha \nonumber\\
&&+\frac{3\lambda a^2 M \beta}{\kappa}-\frac{6\lambda b^2 M\beta}{\kappa}+\frac{3\lambda b^2 N^2 \beta}{\kappa M}-\frac{2\lambda a^3 \gamma}{\kappa}+\frac{2\lambda b^3 N\gamma}{\kappa M}-2a^3\rho(a)\gamma\nonumber\\
&&-2a^3 V(\phi)\gamma+\frac{3\lambda a^2 b\xi}{\kappa}-\frac{2\lambda^2 b^3 \xi}{\kappa}-\frac{\lambda b^3 N^2 \xi}{\kappa M^2}-\frac{6\lambda\beta\dot{b}^2}{M}+\frac{6\lambda b\xi \dot{b}^{2}}{M^{2}}-2a^3 N\alpha\rho'(a)\nonumber\\
&&-2a^3 N\varphi V'(\phi)+\frac{3la^2\alpha\dot{\phi}^2}{N}-\frac{la^3 \gamma\dot{\phi}^2}{N^2}-\frac{12\lambda b\dot{b}\beta_t}{M}+\frac{3\lambda a^2 bM\tau_t}{\kappa}\nonumber\\
&&-\frac{2\lambda^2 b^3 M\tau_t}{\kappa}-\frac{2\lambda a^3 N\tau_t}{\kappa}+\frac{\lambda b^3 N^2 \tau_t}{\kappa M}-2a^3 N\rho(a)\tau_t-2a^3 NV(\phi)\tau_t\nonumber\\
&&+\frac{6\lambda b\dot{b}^2 \tau_{t}}{M}-\frac{la^3\dot{\phi}^2 \tau_t}{N}+\frac{2la^3\dot{\phi}\varphi_t}{N}-\frac{24\lambda b\dot{b}\dot{\phi}\beta_\phi}{M}+\frac{6\lambda a^2 bM\dot{\phi}\tau_\phi}{\kappa}\nonumber\\
&&-\frac{4\lambda^2 b^3 M\dot{\phi}\tau_\phi}{\kappa}
-\frac{4\lambda a^{3}N\dot{\phi}\tau_{\phi}}{\kappa}+\frac{2\lambda b^3 N^2\dot{\phi}\tau_\phi}{\kappa M}-4a^3 N\rho(a)\dot{\phi}\tau_\phi-4a^3 N V(\phi)\dot{\phi}\tau_\phi\nonumber\\
&&+\frac{12\lambda b\dot{b}^2 \dot{\phi}\tau_\phi}{M}
-\frac{2la^3\dot{\phi}^3 \tau_\phi}{N}+\frac{4la^3\dot{\phi}^2\varphi_\phi}{N}-\frac{24\lambda b\dot{b}\dot{M}\beta_M}{M}+\frac{6\lambda a^2 bM\dot{M}\tau_M}{\kappa}\nonumber\\
&&-\frac{4\lambda^2 b^3 M\dot{M}\tau_M}{\kappa}-\frac{4\lambda a^3 N\dot{M}\tau_M}{\kappa}+\frac{2\lambda b^3 N^2 \dot{M}\tau_M}{\kappa M}-4a^3 N\rho(a)\dot{M}\tau_M\nonumber\\
&&-4a^3 NV(\phi)\dot{M}\tau_M+\frac{12\lambda b\dot{b}^2 \dot{M}\tau_M}{M}-\frac{2la^3\dot{M}\dot{\phi}^2\tau_M}{N}+\frac{4la^3 \dot{M}\dot{\phi}\varphi_M}{N}\nonumber\\
&&-\frac{24\lambda b\dot{b}\dot{N}\beta_N}{M}+\frac{6\lambda a^2 bM\dot{N}\tau_N}{\kappa}-\frac{4\lambda^2 b^3 M\dot{N}\tau_N}{\kappa}-\frac{4\lambda a^3 N\dot{N}\tau_N}{\kappa}\nonumber\\
&&+\frac{2\lambda b^3N^2\dot{N}\tau_N}{\kappa M}-4a^3 N\rho(a)\dot{N}\tau_N-4a^3NV(\phi)\dot{N}\tau_N+\frac{12\lambda b\dot{b}^2\dot{N}\tau_N}{M}\nonumber\\
&&-\frac{2la^3\dot{N}\dot{\phi}^2\tau_N}{N}+\frac{4la^3 \dot{N}\dot{\phi}\varphi_N}{N}-\frac{24\lambda b\dot{b}^2\beta_b}{M}+\frac{6\lambda a^2 b M\dot{b}\tau_b}{\kappa}-\frac{4\lambda^2 b^3 M\dot{b}\tau_b}{\kappa}\nonumber\\
&&-\frac{4\lambda a^3 N\dot{b}\tau_b}{\kappa}+\frac{2\lambda b^3 N^2\dot{b}\tau_b}{\kappa M}-4a^3N\rho(a)\dot{b}\tau_b-4a^3NV(\phi)\dot{b}\tau_b+\frac{12\lambda b\dot{b}^3 \tau_b}{M}\nonumber\\
&&-\frac{2la^3 \dot{b}\dot{\phi}^2 \tau_b}{N}+\frac{4la^3\dot{b}\dot{\phi}\varphi_{b}}{N}-\frac{24\lambda b\dot{a}\dot{b}\beta_a}{M}+\frac{6\lambda a^2 b M \dot{a}\tau_a}{\kappa}-\frac{4\lambda^2 b^3 M\dot{a}\tau_a}{\kappa}\nonumber\\
&&-\frac{4\lambda a^3 N\dot{a}\tau_a}{\kappa}+\frac{2\lambda b^3 N^2 \dot{a}\tau_a}{\kappa M}-4a^3 N\rho(a)\dot{a}\tau_a-4a^3 NV(\phi)\dot{a}\tau_a+\frac{12\lambda b\dot{a}\dot{b}^2\tau_a}{M}\nonumber\\
&&-\frac{2la^3\dot{a}\dot{\phi}^2 \tau_a}{N}+\frac{4la^3 \dot{a}\dot{\phi}\varphi_a}{N}=B_t\nonumber
+\dot{a}B_a +\dot{b}B_b +\dot{M}B_M+\dot{N}B_N+\dot{\phi}B_\phi.
\end{eqnarray}
If the Noether symmetry condition $\pounds_{\mathrm{X}_{\rm NGS}}{\mathcal{L}_{\rm EiBI}}=0$ is satisfied, then the function $\Sigma_{0}=\alpha^{i}\frac{\partial \mathcal{L}}{\partial \dot{q}^{i}}$ is a constant of motion \cite{Copoz Ritis96}. This gives
\begin{eqnarray}\label{constant of motion EiBI}
\Sigma_{0,\rm EiBI}&=&\alpha^{i}\frac{\partial \mathcal{L}}{\partial \dot{q}^{i}}=\beta\frac{\partial\mathcal{ L}}{\partial \dot{b}}+\varphi\frac{\partial \mathcal{L}}{\partial \dot{\phi}}=
\beta\Big[  -\frac{12\lambda b\dot{b}}{M} \Big]+\varphi \Big[ \frac{2a^{3}l\dot{\phi}}{N}  \Big],
\end{eqnarray}
where two unknown functions $\beta$ and $\varphi$ will be studied in the next section.
Up to this point, it is worth mentioning a dimension analysis of each variable, i.e. $[\rm dimensionless]=[\lambda]=[l]= [$a$]=[$b$]=[\phi]=[N]=[M];\quad[\alpha]=[\beta]=[\gamma]=[\xi]=[\varphi]=[\tau]=[M_{P}^{-1}];[\kappa]=[M_{P}^{-2}]; \quad[\dot{a}]=[\dot{b}]=[\dot{\phi}]=[M_{p}];\quad [\varphi_{t}]=[\beta_{t}]=[\tau_{t}]=[M_{P}^{-2}]$.

%%%%%%%%%%%%%%%%%%%%%%%%%%%%%%%%%%%%%%%%%%%%%%%%%
\section{Remarks on exact cosmological solutions}
\label{Sols}
%%%%%%%%%%%%%%%%%%%%%%%%%%%%%%%%%%%%%%%%%%
After a separation of monomials, we can quantify the system equations to yield
\begin{eqnarray}
\tau_a &=& \tau_b = \tau_M = \tau_N =\tau_{\phi}=0,\label{tau t}\\ 
\beta_a &=&\beta_M=\beta_N=0, \label{beta aMN}\\
\varphi_a &=&\varphi_M=\varphi_N=0, 
\label{varphi aMN}\\
0&=&-\frac{6\lambda b \beta_{\phi}}{M}+\frac{la^{3}\varphi_{b}}{N},\label{varphi and beta2}\\
0&=&3\alpha-\frac{a\gamma}{N}-a\tau_{t}+4a\varphi_{\phi}
,\label{gamma const}\\
0 &=&  -\beta+\frac{b\xi}{M}+b\tau_{t}-4b\beta_{b},
\label{xi const}\\
B_{a}&=&B_{M}=B_{N}=0 \\
B_b &=& -\frac{12\lambda b\beta_t}{M},\label{B be}\\
B_\phi &=& \frac{2la^{3}\varphi_{t}}{N},\label{B phi}\\
B_t &=& \alpha\bigg[ \frac{6\lambda abM}{\kappa}-\frac{6\lambda a^2 N}{\kappa}-{6a^2 N\rho(a)}-6a^2 NV(\phi) -2a^3 N\rho'(a)\bigg]\label{B t redu}\nonumber\\ &&+\beta\bigg[\frac{3\lambda a^2 M}{\kappa}
-\frac{6\lambda^2 b^2 M}{\kappa}+\frac{3\lambda b^2 N^2}{\kappa M}  \bigg]+\gamma\bigg[-\frac{2\lambda a^3}{\kappa}+\frac{2\lambda b^3 N}{\kappa M}\nonumber\\
&&-2a^3 \rho(a)-2a^3 V(\phi)  \bigg]+\xi \bigg[ \frac{3\lambda a^2 b}{\kappa}-\frac{2\lambda^2 b^3 }{\kappa}-\frac{\lambda b^3 N^2 }{\kappa M^2} \bigg]\nonumber+ \varphi\Big[ -2a^3 NV'(\phi)\Big]\\ &&+\tau_{t}\bigg[\frac{3\lambda a^{2}bM}{\kappa}-\frac{2\lambda^{2}b^{3}M}{\kappa} -\frac{2\lambda a^{3}N}{\kappa}+\frac{\lambda b^{3}N^{2}}{\kappa M} -2a^{3}N\rho(a)-2a^{3}NV(\phi)\bigg].
\end{eqnarray}
%The appearance of the NGS constraint equations as shown in the previous equations helps us to reduce the complexity of the single constraint equation that appeared in the previous section.
From Eq.(\ref{beta aMN}),Eq.(\ref{varphi aMN} ) and Eq.(\ref{varphi and beta2}), one found that
$\beta=\beta(b)$ and $\varphi(\phi).$
If we choose
\begin{eqnarray}\label{beta funt}
\beta(b)={c}_{1}b,\\ \varphi(\phi)={c}_{2}\phi.
\end{eqnarray}
From Eq.(\ref{tau t}), there is only one possibility left with $\tau_{t}\neq 0.$ Therefore the polynomial of $\alpha, \beta,\varphi,\gamma$, and $\xi$ no longer have hold in $\tau(t).$  Proposing the linearity of the relations expressed in Eq.(\ref{gamma const}) and Eq.(\ref{xi const}), we have to set 
 \begin{eqnarray}
  \alpha(a)&=&c_{3}a, \\
  \gamma(N)&=&c_{4}N,\\
  \xi(M)&=&c_{5}M.
 \end{eqnarray}
With this setting, we can solve Eq.(\ref{gamma const}) and  Eq.(\ref{xi const}) to get 
\begin{eqnarray}
\tau(t)&=&(3c_{3}-c_{4}+4c_{2})t+c_{6},\\
\tau(t)&=&(5c_{1}-c_{5})t+c_{7}.
\end{eqnarray}
In order to write $\tau(t)$ in a  single form, we have to set
 $c_{6}=c_{7}$ and $3c_{3}-c_{4}+4c_{2}=5c_{1}-c_{5}.$
%The conditions that allows $\tau(t)=0$ are $c_{6}=c_{7}=0$, $c_{5}=5c_{1}$, and $c_{4}=4c_{2}+3c_{3}.$ This is
% \begin{eqnarray}
% \xi(M)&=&5c_{1}M,\\
% \gamma(N)&=&(4c_{2}+3c_{3})N.
%\end{eqnarray}
There are three equations contributed of gauge function, 
\begin{eqnarray}
B_{b}&=&-\frac{12\lambda b \beta_{t}}{M}=-\frac{12c_{1}\lambda b\dot{b}}{M}, \label{Gau B b}\\ 
B_{\phi}&=& \frac{2la^{3}\varphi_{t}}{N}=\frac{2c_{2}la^{3}\dot{\phi}}{N}, \label{Gau B phi}\\
B_t&=& c_{3}\bigg[ \frac{6\lambda a^{2}bM}{\kappa}-\frac{6\lambda a^3 N}{\kappa}-{6a^3 N\rho(a)}-6a^3 NV(\phi) -2a^3 N\rho'(a)\bigg]\label{B t redu2}\nonumber \\&&+c_{1}\bigg[\frac{3\lambda a^2b M}{\kappa}
-\frac{6\lambda^2 b^3 M}{\kappa}+\frac{3\lambda b^3 N^2}{\kappa M}  \bigg]+c_{4}\bigg[-\frac{2\lambda a^3N}{\kappa}+\frac{2\lambda b^3 N^{2}}{\kappa M}\nonumber\\
&&-2a^3 N\rho(a)-2a^3N V(\phi)  \bigg]+c_{5} \bigg[ \frac{3\lambda a^2 bM}{\kappa}-\frac{2\lambda^2 b^3M }{\kappa}-\frac{\lambda b^3 N^2 }{\kappa M} \bigg]\nonumber\\&&+(3c_{3}-c_{4}+4c_{2})\bigg[ \frac{3\lambda a^{2}bM}{\kappa} -\frac{2\lambda^{2}b^{3}M}{\kappa}-\frac{2\lambda a^{3}N}{\kappa}+\frac{\lambda b^{3}N^{2}}{\kappa M}-2a^{3}N\rho(a)\nonumber \\&&-2a^{3}NV(\phi)\bigg]-2c_{2}a^3 N\phi V'(\phi),
\end{eqnarray}
where we use $\beta_{t}=\frac{\partial \beta}{\partial b}\frac{db}{dt}=c_{1}\dot{b}$ and $\varphi_{t}=\frac{\partial \varphi}{\partial \phi}\frac{d\phi}{dt}=c_{2}\dot{\phi}$ to get Eq.(\ref{Gau B b}) and (\ref{Gau B phi}). We hence
expect that the rest of the expression for boundary term, i.e. $-\frac{12c_{1}\lambda b\dot{b}}{M},$ may relate with  $B_{t}$. 
From Eq.(\ref{varphi and beta2}), Eq.(\ref{B phi}) and Eq.(\ref{B be}), it is easy to see that
\begin{eqnarray}
\frac{6\lambda b}{M}=-\frac{B_{b}}{2\beta_{t}},\quad
\frac{la^{3}}{N}=\frac{B_{\phi}}{2\varphi_{t}}.
\end{eqnarray}
This gives the following relation,
\begin{eqnarray}
\frac{6\lambda b\beta_{\phi}}{M}=\frac{la^{3}\varphi_{b}}{N}, \\
-\frac{B_{b}}{2\beta_{t}}\beta_{\phi}=\frac{B_{\phi}}{2\varphi_{t}}\varphi_{b},
\end{eqnarray}
This confirms again that $\beta_{\phi}=\varphi_{b}=0$, but keeps  $\beta_{t}\neq 0$ and $\varphi_{t} \neq 0$. That also means that $B_{b}\neq 0$ and $B_{\phi}\neq 0.$
The boundary term can be also partly derived from Eq.(\ref{Gau B b}) and Eq.(\ref{Gau B phi}), that is
\begin{eqnarray}
B_{(b,\phi)}=-\frac{6c_{1}\lambda b^{2}\dot{b}}{M}+\frac{2c_{2}a^{3}l\phi\dot{\phi}}{N}.
\end{eqnarray}
 It is worth noting that  $c_{1}$ is just an arbitrary constant and we can redefine it by replacing $c_{1}\rightarrow 2c_{1}$. Interestingly, this is exactly matched with the constant of motions of EiBI gravity by  this setting.
\begin{eqnarray}
\Sigma_{{0},\rm {EiBI}}&=&-\frac{12c_{1}\lambda b^{2}\dot{b}}{M}+\frac{2c_{2}a^{3}l\phi\dot{\phi}}{N}.
\end{eqnarray}
The relation between the the constant of motion and the boundary term has shown in \cite{Kurt-Fund}. 
Eq.(\ref{B t redu2}) can be rewritten as
\begin{eqnarray}
B_{t}&=&\frac{\lambda a^{3}N}{\kappa}\bigg[-12c_{3}-8c_{2}\bigg] + a^{3}N\rho(a)\bigg[ -12c_{3}-8c_{2}\bigg]+a^{2}NV(\phi)\bigg[-12c_{3}-8c_{2}\bigg]\nonumber\\ && +\frac{\lambda a^{2}bM}{\kappa}\bigg[3c_{1}+12c_{2}+15c_{3}-3c_{4}+3c_{5}\bigg]\nonumber\\&&
+\frac{\lambda^{2}b^{3}M}{\kappa}\bigg[ -6c_{1}-8c_{2}-6c_{3}+2c_{4}-2c_{5} \bigg]\nonumber\\&&
+\frac{\lambda b^{3}N^{2}}{\kappa M}\bigg[ 3c_{1}+4c_{2}+3c_{3}+c{4}-c_{5}\bigg]\nonumber \\&&-2a^{3}N\bigg[c_{3}\rho'(a)+c_{2}V'(\phi)\bigg].
\end{eqnarray}
where it is very naturally to set  $c_{4}=c_{5}$ and $c_{3}=-\frac{2}{3}c_{2}.$  It is quite remarkable that most terms of $B_{t}$ become zero with the simple setting, i.e.
\begin{eqnarray}
B_{t}&=&
\frac{\lambda a^{2}bM}{\kappa}\bigg[3c_{1}+2c_{2}\bigg]
+\frac{\lambda^{2}b^{3}M}{\kappa}\bigg[3c_{1}+2c_{2}\bigg]
+\frac{\lambda b^{3}N^{2}}{\kappa M}\bigg[3c_{1}+2c_{2} \bigg]\nonumber\\&&-2c_{2}a^{3}N\bigg[-\frac{2}{3}\rho'(a)+V'(\phi)\bigg].
\end{eqnarray}
If we further set that  $c_{1}=-\frac{2}{3}c_{2}=-\frac{2}{3}(-\frac{3}{2})c_{3}=c_{3},$ it is worth seeing that
\begin{eqnarray}\label{Guage B}
B_{t}
%\equiv-\frac{12c_{1}\lambda b\dot{b}}{M}
&=&c_{1}a^{2}N\bigg[ -2\rho'(a)+3V'(\phi) \bigg],\nonumber\\&=& 6c_{1}a^{2}N\bigg[ \frac{\rho_{m}(a)+P_{m}(a)}{a}+V'(\phi)\bigg],
\end{eqnarray}
where the continuity equation, $\rho'(a)=-3(\frac{(\rho_{m}+P_{m})}{a}),$ has been used to get Eq.(\ref{Guage B}). Due to the appearance of $c_{1}$ on the right-hand side of Eq.(\ref{Guage B}), it is reasonable to set $B_{t} \equiv -\frac{12c_{1}\lambda b\dot{b}}{M}.$ This gives the relation between two scale factors, $b(t)$ and $a(t)$, as shown below \begin{eqnarray}
b\dot{b}=-\frac{3a^{2}N(t)M(t)}{1+\kappa \Lambda}\bigg[ \frac{\rho_{m}(a)+P_{m}(a)}{a}+V'(\phi)\bigg]\label{b and a rel}.
\end{eqnarray}
To explain the expansion universe at late time, the exponential potential, i.e. $V(\phi)=V_{0}e^{-\phi}$ is suitable for the model of gravity than
the power law potential, $V(\phi)=V_{0}\phi^{2}.$ For the exponential potential, this gives
\begin{eqnarray}
b\dot{b}=-\frac{3a^{2}N(t)M(t)}{1+\kappa \Lambda}\bigg[ \frac{\rho_{m}(a)+P_{m}(a)}{a}-V_{0}e^{-\phi}\bigg]\label{b and a rel exponent}>0.
\end{eqnarray}
whereas 
\begin{eqnarray}
b\dot{b}=-\frac{3a^{2}N(t)M(t)}{1+\kappa \Lambda}\bigg[ \frac{\rho_{m}(a)+P_{m}(a)}{a}+2V_{0}\phi(t)\bigg]<0.\label{b and a rel power law}
\end{eqnarray}
for the power laws potential. Here we interest to examine further by setting $V_{0}e^{-\phi}\simeq V_{0}(1-\phi(t))$ where $\phi(t)\ll 1$ and using the limit ranged of $\kappa \Lambda$ is $1.12\times 10^{-4} \lesssim \kappa \Lambda \lesssim 2.10\times 10^{-3}$, $M= \sqrt{1+\kappa \Lambda},N(t)=1$, $\frac{a^{2}}{b^{2}}=\frac{1}{1+\kappa \Lambda}$ as shown in \cite{NK2020}. This gives the de Sitter solution in EiBI gravity model in the eye of Noether gauge symmetry,
\begin{eqnarray}
\frac{\dot{b}}{b}\simeq \frac{3V_{0}}{\sqrt{1+\kappa\Lambda}},\quad
b(t)=e^{\frac{3V_{0}t}{\sqrt{1+\kappa \Lambda}}}.
\end{eqnarray}
Clearly from Eq.(\ref{b and a rel exponent}), if there is no  contribution of matter fields, i.e. $\rho_{m}=0,$ the scalar field $\phi(t)\rightarrow 0$, this allows $H_{b}=\frac{\dot{b}}{b}\rightarrow {\rm const}$. This is the de Sitter phase of EiBI Universe.

\section{Conclusion  \label{summary}}

%%%%%%%%%%%%%%%%%%%%%%%%%%%%%%%

We revisited a formal framework of the Eddington-inspired Born-Infeld (EiBI) theory of gravity and derived the point-like Lagrangian for underlying theory based on the use of Noether gauge symmetries (NGS). A Hessian matrix and quantify Euler-Lagrange equations of EiBI universe have been explicitly quantified. We also discussed the NGS approach for the Eddington-inspired Born-Infeld theory and comment on exact cosmological solutions.

We end this work by providing some remarks. As expected, the NGS method can simplify the complication of constraint equations and also helps us to simplify further the gauge function equations with the linear forms of $\beta(b),\varphi(\phi), \alpha(a),\gamma(N), \xi(M)$ and $\tau(t).$  By assuming the equality of $B_{t}$ and the constant of motion, the two scale factors $a(t)$ and $b(t)$ are correlated through the matter fields and the scalar field. Interestingly, we show that there exists the de Sitter solution in this gravity model.
%The linear forms of $\alpha(a),\beta(b),\varphi(\phi)$ and $\tau(t),$ that are compatible with the structural of the EiBI gravity theory, can reduce the complexity of the constraint terms. Two cosmic variables $ b(t)$ and $\phi(t)$ are wait to be examined via the relation between the gauge function constraints and the vanishing of three canonical momenta, i.e. $p_{a},p_{N}$, and $p_{M}$.

%%%%%%%%%%%%%%%%%%%%%%%%%%%%%%%%%
%We end this work by providing some remarks. As it is expected, the NGS method can be used to simplify the complication of the constraint equations and provides the linear forms of $\beta(b),\varphi(\phi), \alpha(a),\gamma(N), \xi(M)$ and $\tau(t)$  help us to simplify further the gauge functions equations. Future work is planned to  interpret the physical meaning of $B_{t}=6c_{1}a^{2}N\bigg[ \frac{\rho_{m}(a)+P_{m}(a)}{a}+V_{0}\phi(t)\bigg].$ 

%and examine if this term relate to  that expects to deeply relate to  $B_{t}.
%The relation between the gauge function constraints can evaluate the relation among two scale factors, matter fields and scalar field. 

%------------------------------------
\subsection*{Acknowledgments}
T. Kanesom is financially supported by the Science Achievement Scholarship of Thailand (SAST) and Graduate School and the Department of Physics, Prince of Songkla University. P. Channuie acknowledged the Mid-Cereer Research Grant 2020 from National Research Council of Thailand under a contract No. NFS6400117.
%------------------------------------

\end{document}